\documentclass[twocolumn,showpacs,preprintnumbers,amsmath,amssymb,superscriptaddress]{revtex4}
\usepackage{graphicx}% Include figure files
\usepackage{dcolumn}% Align table columns on decimal point
\usepackage{color}
\usepackage{bm}% bold math
\usepackage[tight]{units}
\usepackage[thinspace,amssymb]{SIunits}
 \newcommand{\Ce}{CeRu$_4$Sn$_6$} \newcommand{\La}{LaRu$_4$Sn$_6$}
\newcommand{\Y}{YRu$_4$Sn$_6$} % \nofiles

\begin{document} \preprint{APS/123-QED}

 \title{CeRu$_4$Sn$_6$: heavy fermions emerging from a Kondo-insulating
state}% Force line breaks with \\

 \author{E.M.~Br\"uning}\email{bruening@cpfs.mpg.de} \author{M.~Brando}
\author{M.~Baenitz}
\affiliation{Max Planck Institute for Chemical
Physics of Solids\\ N\"othnitzer Stra\ss e 40, 01187 Dresden, Germany}
\author{A.~Bentien}
\affiliation{Max Planck Institute for Chemical
Physics of Solids\\ N\"othnitzer Stra\ss e 40, 01187 Dresden, Germany}
\author{A.M.~ Strydom}
\affiliation{Physics Department, University of
  Johannesburg, South Africa}
\author{R.E.~Walstedt} \affiliation{Physics
  Department, University of Michigan, Ann Arbor, USA}
\author{F.~Steglich}
\affiliation{Max Planck Institute for Chemical Physics of Solids\\
  N\"othnitzer Stra\ss e 40, 01187 Dresden, Germany}
\date{\today}% It is always \today, today, % but any date may be

 %---------------------------------------Abstract------------------------------------------
 \begin{abstract} The combination of low-temperature specific-heat and
nuclear-magnetic-resonance (NMR) measurements reveals important
information of the ground-state properties of \Ce, which has been
proposed as a rare example of a tetragonal Kondo-insulator (KI). The NMR
spin-lattice-relaxation rate $1/T_{1}$ deviates from the Korringa law
below \unit{100}{\kelvin} signaling the onset of an energy gap $\Delta
E_{g1}/k_B \approx \unit{30}{\kelvin}$. This gap is stable against
magnetic fields up to $\unit{10}{\tesla}$. Below
\unit{10}{\kelvin}, however, unusual low-energy excitations of in-gap
states are observed, which depend strongly on the field $H$. The
specific heat $C$ detects these excitations in the form of an enhanced
Sommerfeld coefficient $\gamma = C(T)/T$: In zero field, $\gamma$
increases steeply below \unit{5}{\kelvin}, reaching a maximum at
\unit{0.1}{\kelvin}, and then saturates at $\gamma \approx
\unit{0.6}{\joule\per\mole\kelvin\squared}$. This maximum is shifted to
higher temperatures with increasing field suggesting a residual density
of states at the Fermi level developing a spin gap $\Delta E_{g2}$. A
simple model, based on two narrow quasiparticle bands located at the
Fermi level - which cross the Fermi level in zero field at 0.022
states/meV f.u. - can account qualitatively as well as quantitatively
for the measured observables. In particular, it is demonstrated that
fitting our data of both specific heat and NMR to the model,
incorporating a Ce magnetic moment of $\mu = \Delta E_{g1} /
\mu_{0}H\approx 1~\mu_{B}$, leads to the prediction of the field
dependence of the gap.\\ Our measurements rule out the presence of a
quantum critical point as the origin for the enhanced $\gamma$ in \Ce\
and suggest that this arises rather from correlated, residual in-gap
states at the Fermi level. This work provides a fundamental route for
future investigations into the phenomenon of narrow-gap formation in the
strongly correlated class of systems.
 \end{abstract}
 %-----------------------------------------------------------------------------------------
  % PACS, the
 %Physics and Astronomy % Classification Scheme. %\keywords{Suggested
 %keywords}%Use showkeys class option if keyword %display desired
 \maketitle
 %------------------------------------Introduction-----------------------------------------
 \section{Introduction}

 Among the wide variety of electronic and magnetic ground states played
out by the variable hybridization between a local magnetic moment and
degenerate conduction-electron states, the formation of a semiconducting
ground state remains enigmatic. The simple picture of a single
half-filled conduction band mixing with one magnetic level per site to
produce the Kondo insulating (or heavy-fermion semiconducting) state
gives an appealingly simple description for this class of materials. The
body of experimental evidence is not so unequivocal however, and even
the involvement of hybridization which drives the Kondo interaction and
supposedly re\-ali\-zes a semiconducting state in theses systems, has been
brought into question.

Kondo insulator (KI) materials (e.g., see \cite{Jaime2000} and
references therein) are Kondo lattice (KL) systems which exhibit
semiconducting behavior below a certain temperature $T_{g}$ at which an
energy gap opens at the Fermi level~\cite{Aeppli1992,Riseborough2000}.
Although numerous investigations have been performed on Kondo insulator
compounds, where the behavior of the gap has been observed under the
effect of external parameters like pressure~\cite{Gabani2003} or
magnetic field~\cite{Jaime2000}, the intrinsic conditions under which a
KL transforms into a KI are still not clear. A puzzling phenomenon that
complicates the concept of a hybridization gap~\cite{Aeppli1992}, is the
fact that an electrically insulating ground state as $T \to 0$ is rather
uncommon. In this group of KI systems a small residual carrier
concentration and hence a finite electrical conductivity at $T \to 0 $
appears to be generic, as observed at low temperature
\cite{Strydom2005}.

 The general physical properties of \Ce\ are attributed to the formation
of the Kondo insulating state in this compound
\cite{Das1992,Strydom2005}. The residual carrier levels in \Ce\ have a
magnetic origin and yet muon spin relaxation ($\mu$SR) experiments
\cite{Strydom2007}, a probe which is exceedingly sensitive to magnetic
cooperation phenomena, prove the ground state of \Ce\ to be free from
long-range or even short-range ordering. \Ce\ belongs to a family of
rare-earth R-Ru$_{4}$Sn$_{6}$ ternary stannides
\cite{Poettgen1997,Koch2008}. Together with U$_{2}$Ru$_{2}$Sn, it is a
rare occurrence of KI having tetragonal crystal structure
(crystallographic space group $I\bar{4}2m$). Evidence of the onset of an
energy gap at about \unit{30}{\kelvin} has been found in the temperature
dependence of several quantities: electrical resistivity ($\rho$) and
thermal conductivity ($\kappa$) \cite{Das1992,Strydom2005},
spin-lattice-relaxation rate ($1/T_{1}$) in nuclear-magnetic-resonance
(NMR) experiments~\cite{Bruening2007} and thermopower
($S$)~\cite{Strydom2005}.

\begin{figure}[!ht] \centering
  \includegraphics[width=0.97\columnwidth]{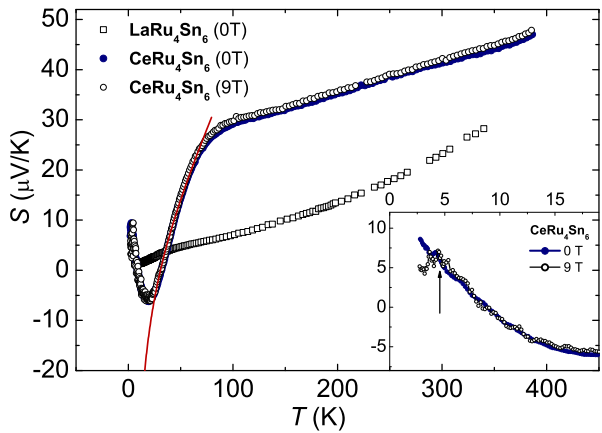}
  \caption{(Color online) Temperature dependence of the thermopower
    measured in 0 and \unit{9}{\tesla}. The line shows a fit according
    to an energy gap-derived contribution $S(T)_{gap}\propto\Delta
    E_{g1}/k_B T$ (see text). Inset: Low-temperature $S(T)$ data at 0
    and \unit{9}{\tesla}. The arrow indicates the maximum observed at \unit{9}{\tesla}.}
  \label{fig:thermo}
\end{figure}

The $T$-dependent thermopower of \Ce\ and
\La\ is shown in Fig.~\ref{fig:thermo}.
It was measured in zero field and at \unit{9}{\tesla} in the temperature
range between \unit{2} and \unit{375}{\kelvin}. In the measured
temperature range there is no indication of peaks which can be
attributed to a crystalline electric field (CEF) splitting. This is
consistent with inelastic neutron scattering
experiments~\cite{Adroja2006} from which a resonance of magnetic origin
at about \unit{30}{\milli\electronvolt} $\approx \unit{350}{\kelvin}$ is
the only evidence of CEF effects in \Ce\ in this temperature range. For
\La~(zero field), $S$ shows a linear temperature dependence for $T \to
0$ which is characteristic of simple metals \cite{Blatt1976}. In fact, $S$ can be
obtained from the thermoelectric potential by applying a temperature
gradient, and in metals $S(T)$ commonly tracks the electronic density of
states at the Fermi energy:
\begin{equation}
  \label{eq:1} S \propto T \left(\frac{\partial \ln N(E)}{\partial E}
\right)_{E=E_F}.
\end{equation}

In a rigid band model where $N(E)$ is temperature independent, $S$ is
therefore proportional to $T$. The linear relation could also be observed
in the thermopower of \Ce\ above $\simeq \unit{100}{\kelvin}$, even
though the absolute values are enhanced, compared to those of \La. Above
\unit{50}{\kelvin}, the thermopower can be ascribed to a conduction
band in a semi-metal \cite{Blatt1976}. Here, $S(T)$ can be expressed as
a sum of a linear contribution and a contribution $S(T)_{gap} \propto
\Delta E_{g1}/k_{B}T$ ($\Delta E_{g1}/k_{B} = \unit{36}{\kelvin}$)
\cite{Strydom2005} (line in Fig~\ref{fig:thermo}). The size of the
energy gap is consistent with that found in resistivity
measurements~\cite{Das1992}. Towards lower temperatures, $S$ strongly
decreases and reaches a maximum negative value at $T \simeq
\unit{22}{\kelvin}$. This is characteristic of correlated semimetals
with a residual density of states within an energy gap, as is also found
for example in the Kondo insulator SmB$_6$~\cite{Guo2005}. At
temperatures lower than the \unit{22}{\kelvin} extremum, $S(T)$ turns
toward $S = 0$ which is achieved near $\unit{8}{\kelvin}$, but further
cooling sees an increase in $S$ towards positive values. % The peak in
% $S(T)$ found at $T = \unit{3}{\kelvin}$ is expected to be followed by a
% steep decrease at lower temperatures since
Since the third law of thermodynamics predicts $S = 0$ at $T = 0$, a
maximum in $S(T)$ is expected towards lower temperatures
\cite{Zlatic2007}. The temperature range between 2 and
\unit{15}{\kelvin} is shown in the inset of Fig.~\ref{fig:thermo}. Here,
an applied field of $\mu_{0}H = \unit{9}{\tesla}$ shifts the maximum
expected at low $T$  in $S(T)$ to around \unit{5}{\kelvin} and indeed projects the
thermopower back towards zero as $T\to 0$. This observed field
dependence likely suggests further low-lying phenomena such as a small
gapped structure within the narrow f-band. The sensitive nature of
$S(T)$ to fields is in accord with the magnetic nature of the in-gap
states. This interpretation is also supportive of former heat-capacity
results performed in magnetic field for temperatures larger than
\unit{0.3}{\kelvin}, where a maximum is observed in $C/T$ vs. $T$ which
shifts to higher temperature with
increasing $H$~\cite{Strydom2005}.\\
\begin{figure}[!ht] \centering
\includegraphics[width=0.97\columnwidth]{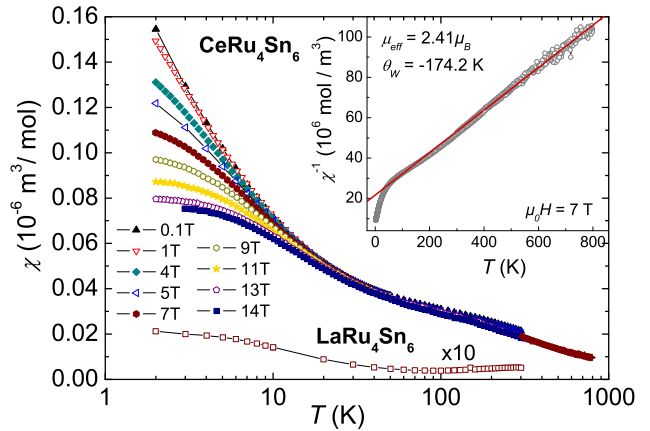}
  \caption{(Color online) Magnetic susceptibility
of \Ce\ in a semi-log scale, showing the field-dependence that develops below
\unit{20}{\kelvin} (main panel) and the local-moment magnetic character
inferred from the high-temperature data (inset). For comparison, the
very small and weakly temperature-dependent $\chi$ of non-magnetic \La\
is also shown.}
  \label{fig:susc}
\end{figure}
The concept of magnetic residual in-gap states is proven by the uniform
susceptibility $\chi(T,H)=M(T)/H$ shown in Fig.~\ref{fig:susc}. The
results portray the paramagnetic state with a strong field dependence of
the susceptibility developing below \unit{20}{\kelvin}. The inverse
susceptibility follows a Curie-Weiss law down to \unit{50}{\kelvin},
without any sign of CEF splitting. From the slope we determine an
effective moment of $\mu_{\rm eff} = \unit{2.41}{\mu_{B}/Ce}$ which is
close to the free Ce$^{3+}$ moment ($\mu_{\rm eff} =
\unit{2.54}{\mu_{B}/Ce}$), indicating a well localized Ce moment at high
temperatures. Below \unit{50}{\kelvin}, a strong deviation from the
Curie-Weiss law is observed and it is associated with the opening of a
gap. Considering the involvment of Kondo physics in \Ce\
\cite{Strydom2005}, a partial compensation of the local moments due to
on-site moment screening may also be involved in the temperature
evolution of $\chi(T)$ at intermediate temperatures. However, in the
observed $\chi(T)$ of \Ce\ the opening of the energy gap at about 30~K
does not appear to cause the typical KI-like \emph{decrease} in
$\chi(T)$ below a gap-derived maximum. In the KI class of systems, this
feature originates from the stable local-moment susceptibility at high
temperatures which turns into a demagnetized state at low temperature
through severe hybridization with the degenerate conduction band. In the
present case we rather observe an increase of $\chi$ towards low
temperatures. On cooling, no significant field dependence of $\chi$
could be resolved down to \unit{20}{\kelvin}. Below this temperature the
behavior changes. With increasing magnetic field the susceptibility
tends to level off at a rather small constant value as visible in the
\unit{14}{\tesla} curve. This behavior confirms the presence of a
residual density of states within an energy gap. In contrast to that,
the structural homologue \La~shows a nearly temperature independent
Pauli susceptibility as it is expected for a nonmagnetic metal.

\begin{figure}[!t] \centering
\includegraphics[width=0.97\columnwidth]{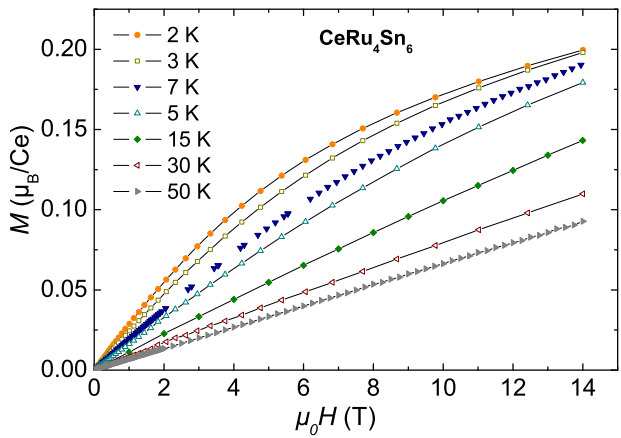}
\caption{(Color online) Magnetization isotherms of CeRu$_{4 }$Sn$_{6}$.} %Lines are guides to the eye.
  \label{fig:Magn}
\end{figure}

The partial density of states accessible to a magnetic field at
\unit{2}{\kelvin} can be also seen in the field dependence of the
magnetization $M$, shown in Fig~\ref{fig:Magn}. At \unit{2}{\kelvin} and
\unit{14}{\tesla} a value of only $M \simeq \unit{0.2}{\mu_{B}/Ce}$ is
reached which corresponds to only about \unit{9.3}{\%} of the full $gJ$
complement of the Ce$^{3+}$ moment. Consequently, no saturation could be
observed up to \unit{14}{\tesla}.

% This might be ascribed to the
% partial compensation of the moment due to the Kondo effect.\\
In this article, we focus essentially on the low-temperature properties of \Ce\
for which the  opening of the \unit{30}{\kelvin} energy gap is a key
ingredient. We demonstrate how two strongly-correlated (narrow)
quasiparticle bands, located at the Fermi level, can account
qualitatively and quantitatively for all properties observed in this
material. Although rather exotic, it is our contention that the scenario
played out in the ground state of \Ce\ has features which make the
proposed model more generally applicable to similar systems in the class
of correlated materials.\\
% Our rather exotic ground state scenario can  be easily %tested with
% band-structu calculations or neutron scattering experiments in magnetic field.\\ 
% In this work we present detailed results of specific heat $C_p(T)$ measurements on \Ce\, to very
% low temperatures and in applied magnetic fields in order to address
% issues of the KI ground state. The magnetic entropy and the correlated
% state are shown to be highly susceptible to applied fields and we
% demonstrate how the energy gap responds to fields. The electronically
% correlated state of the Ce ions are accessible through the hyperfine
% field and the spin-lattice relaxation rate is shown to be a sensitive
% probe of energy gap dynamics.

%------------------------------------Experimental-----------------------------------------

\section{Experimental}

All measurements shown in this article have been carried out on sample
material originating from the same batch. The polycrystalline sample was
synthesized in an arc furnace followed by subsequent phase purifying
heat treatment. The details of the preparation are reported elsewhere
\cite{Strydom2005}. Physical properties, magnetization and thermopower
measurements where carried out in a standard Quantum Design PPMS/MPMS in
the temperature range from 1.8 to \unit{300}{\kelvin}.
The MPMS high-temperature option was used for magnetization measurements
up to \unit{800}{\kelvin}. NMR measurements were obtained in a
temperature range between 2 and \unit{250}{\kelvin} in a $^{4}$He cryostat (Janis) and
with a commercial pulsed NMR spectrometer (TecMag) using the field sweep
method. The field-sweep NMR spectra were performed with different fields
and frequencies, respectively. For the spin-lattice-relaxation
measurements, a saturation-recovery sequence was applied. The heat
capacity has been measured in a dilution refrigerator with the
compensated heat-pulse method~\cite{Wilhelm2004} from \unit{4}{\kelvin} down to 
\unit{0.065}{\kelvin} and in static applied fields up to \unit{10}{\tesla}. 
%
%------------------------------------Results&Discussion-----------------------------------
%
\section{Results}
\subsection{Specific heat}
\label{sec:specific-heat} Field-dependent investigations of the specific
heat $C$ of \La\ and \Ce\ in the temperature range between \unit{0.4} and
\unit{20}{\kelvin} have been reported in Ref.~\cite{Strydom2005}.
\begin{figure}[!ht] \centering
\includegraphics[width=0.97\columnwidth]{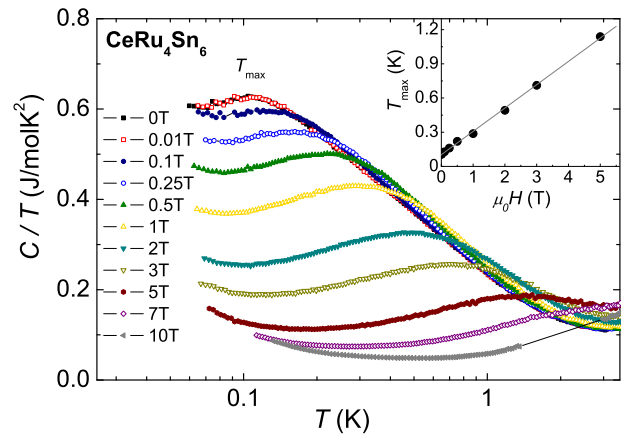}
  \caption{(Color online) Sequence of isofield plots of the low-temperature
 specific heat of CeRu$_{4}$Sn$_{6}$. Inset: Field variation of the
temperature $T_{\rm max}$ where a maximum is achieved in $C(T,H)$, after
subtraction of the nuclear contributions, see text.}
  \label{fig:CRSCpT}
\end{figure}
In zero field, a logarithmic increase of the Sommerfeld coefficient
$\gamma = C/T$ versus $T$ extended, on cooling, over a decade in temperature has been
observed. Non-Fermi-liquid behavior such as
this, similar to what is often seen in strongly correlated systems, for
instance in proximity to a continuous quantum phase transition, has been a subject
of considerable interest. In \Ce\ this particular aspect of the ground
state has been addressed using muon spin relaxation ($\mu$SR) studies
which portrayed an interpretation in
terms of low-temperature quantum fluctuations~\cite{Strydom2007}.

Detailed low-temperature specific heat results are shown in
Fig.~\ref{fig:CRSCpT}. In the whole temperature range, no sign of a
phase transition - magnetic ordering or superconductivity - could be
found in good agreement with $\mu$SR measurements~\cite{Strydom2007}. At
the lowest temperatures a weak increase of $C/T$ with decreasing $T$ is
attributed to a nuclear Schottky contribution. At temperatures
sufficiently larger relative to the nuclear splitting energy, the
high-temperature tail of a Schottky excitation is expected which should follow $C \sim
1/T^{2}$. In the present case, such a contribution likely arises from
$^{119}$Sn and $^{117}$Sn isotopes and $^{99}$Ru and $^{101}$Ru
isotopes. The nuclear Schottky contribution furthermore commonly scales
with the magnetic field as $H^{2}$, as it does in our case. This low-temperature
nuclear contribution is subtracted (Fig.~\ref{fig:CRSCpT}) and is of no further consequence to
the interpretation of our data. 

In zero field, $\gamma(T)$ increases with decreasing temperature, leveling off at a very large value of
\unit{0.6}{\joule\per\kelvin\squared\mole}. In view of the evidence of
absence of any cooperative effects in \Ce\ we attribute this enhanced
specific heat to the formation of heavy electronic quasiparticles. The
Sommerfeld-Wilson ratio, $R_W=\pi^2 k_B^2 \chi(0)/\mu_0\mu_{\rm
  eff}^2\gamma(0)$ ($k_{B}$: Boltzmann-constant) is often used to assess the origin of an enhanced
ground-state magnetic susceptibility and electronic specific heat in
correlated electron systems, both of which are directly related to the
electronic density of states. For a spin $s = 1/2$ Kondo impurity one expects
$R_W=2$ as $T\to 0 $ \cite{wilson1975}. Using $C(T)/T$ and $\chi(T)$
values at $T=\unit{2}{\kelvin}$ in $\mu_0 H= \unit{0.1}{\tesla}$ from
Fig.~\ref{fig:CRSCpT} and Fig.~\ref{fig:susc}, we obtain $R_W\approx 3$.
An enhancement in the susceptibility over the specific heat
typically indicates the presence of ferromagnetic
correlations \cite{Nakatsuji2008}.

As seen in Fig.~\ref{fig:CRSCpT} a local maximum develops at $T_{\rm
  max} = \unit{0.1}{\kelvin}$ which shifts towards higher temperatures with
increasing field. In the inset of Fig.~\ref{fig:CRSCpT}, $T_{\rm max}$
vs. $\mu_0 H$ is plotted and a well defined linear-in-$H$ dependence of
$T_{\rm max}$ is observed. This presumably indicates Zeeman splitting of
degenerate spin states by the applied magnetic field~\cite{Desgranges1982}.\\ From
these measurements we can now conclude that: i) the behavior of $\gamma(T)$
below \unit{5}{\kelvin} is not related to the presence of a quantum
phase transition, but is probably a consequence of correlated in-gap
states, and ii) a presumed degeneracy is lifted in applied
magnetic fields, causing a maximum in $\gamma(T)$.

\subsection{Remarks about the entropy}

The tetragonal crystal structure of \Ce\ would imply a CEF-derived Kramers doublet as
a ground state for the magnetic Ce$^{3+}$ ion with an entropy contribution of
$R\ln 2 = \unit{5.76}{\joule\per\mole\kelvin}$.
\begin{figure}[!ht] \centering
\includegraphics[width=0.97\columnwidth]{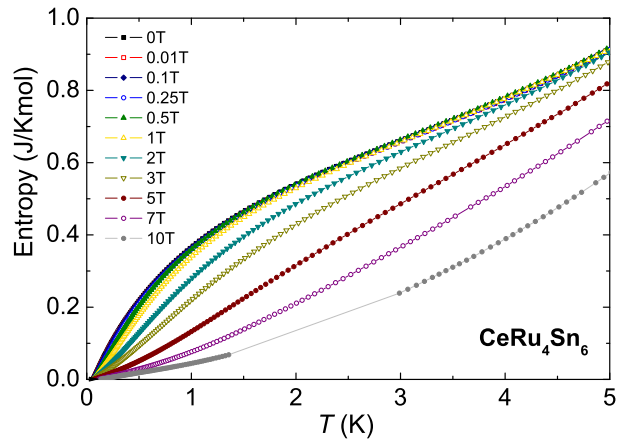}
  \caption{(Color online) Temperature dependent entropy of \Ce\
obtained at various fields up to \unit{10}{\tesla}.}
  \label{fig:entropy}
\end{figure}
In the absence of significant CEF splitting, deduced from our
thermopower and suceptibility measurements above $T\approx \Delta
E_{g1}/k_B = \unit{30}{\kelvin}$ up to \unit{800}{\kelvin}, one may
expect the entropy of the non-degenerate $J = 5/2$ state of Ce$^{3+}$,
$R \ln 6$, to be quenched below \unit{30}{\kelvin} where the gap opens
and the Kondo effect starts to screen the Ce moments. However, in
inelastic neutron scattering experiments a $\boldsymbol{q}$-independent
peak at about \unit{30}{\milli\electronvolt} $\approx
\unit{350}{\kelvin}$ has been detected which likely signals a high-lying
CEF-derived doublet~\cite{Strydom2010}. This could explain the small
discrepancy between the effective moment found in the susceptibility
(cf. Fig.~\ref{fig:susc}) and that of the full Ce$^{3+}$ free-ion
moment. Nonetheless, this still leaves a magnetic entropy amounting to
$R\ln4$ for the ground state. The entropy of CeRu$_{4}$Sn$_{6}$, after
having subtracted the nuclear but not the phonon contribution, is
plotted in Fig.~\ref{fig:entropy}: Surprisingly, at \unit{5}{\kelvin},
less than \unit{10}{\%} of $R\ln2$ is recovered in zero field. Applied
magnetic fields shift the entropy towards higher temperatures. We
performed high-resolution measurements of two non-magnetic isostructural
compounds, \La\ and \Y, with a view to subtracting the phononic
contribution and check how much entropy is recovered at
\unit{30}{\kelvin}. However, as was mentioned in
Ref.~\cite{Strydom2005}, caution has to be excercised in this procedure,
and the La compound may not be considered as a proper phononic analogue
for the Ce one in this class of materials. A similar conclusion was
drawn, for instance, on the KI CeRhSn ~\cite{Slebarski2004}. There are
two main reasons for that: (i) the non-magnetic compounds \La\ and \Y\
are metallic at low temperature whereas \Ce\ is not, and (ii) all these
materials form in a so-called cage structure which exhibits not only
acoustic phonon modes but also optical ones due to the anharmonic motion
of caged-up atoms in the cage framework. The Einstein frequencies of
these modes depend on the lattice constants, and they will give
generally different contributions to the specific heat of \La\ and
CeRu$_{4}$Sn$_{6}$. Indeed, the measured $C(T)$ curves of \Ce\, and \La\
(or \Y ) were found to cross each other below \unit{10}{\kelvin}.
Attemps to subtract the electronic specific heat of \La\ from that of
\Ce\ were inconclusive for understanding at which temperature at least an
amount of entropy close to $R\ln 2$ is recovered. Since the Ce magnetic
moment at \unit{0.06}{\kelvin} is small and thus only a tiny amount of
entropy is left to be released below this temperature, it follows that
at least \unit{90}{\%} of $R\ln2$ has to be distributed between
\unit{5}{\kelvin} and \unit{30}{\kelvin}, in agreement with the
assumption that around \unit{30}{\kelvin} a gap
opens.\\ % In support of the this is the fact that measurements of
% optical reflectivity on \Ce~do not show any signature of a charge gap
% above \unit{100}{\kelvin} \cite{Guritanu2010}.

Considering NMR as a sensitive microscopic probe for the formation of an
energy gap, as well as the involvement of correlated states and
their evolution in a magnetic field, we proceed to describe in the next
section our $^{119}$Sn-NMR results.
\subsection{$^{119}$Sn NMR Spectroscopy}

\label{sec:nmr} 
\begin{figure}[!ht] \centering
  \includegraphics[width=0.85\columnwidth]{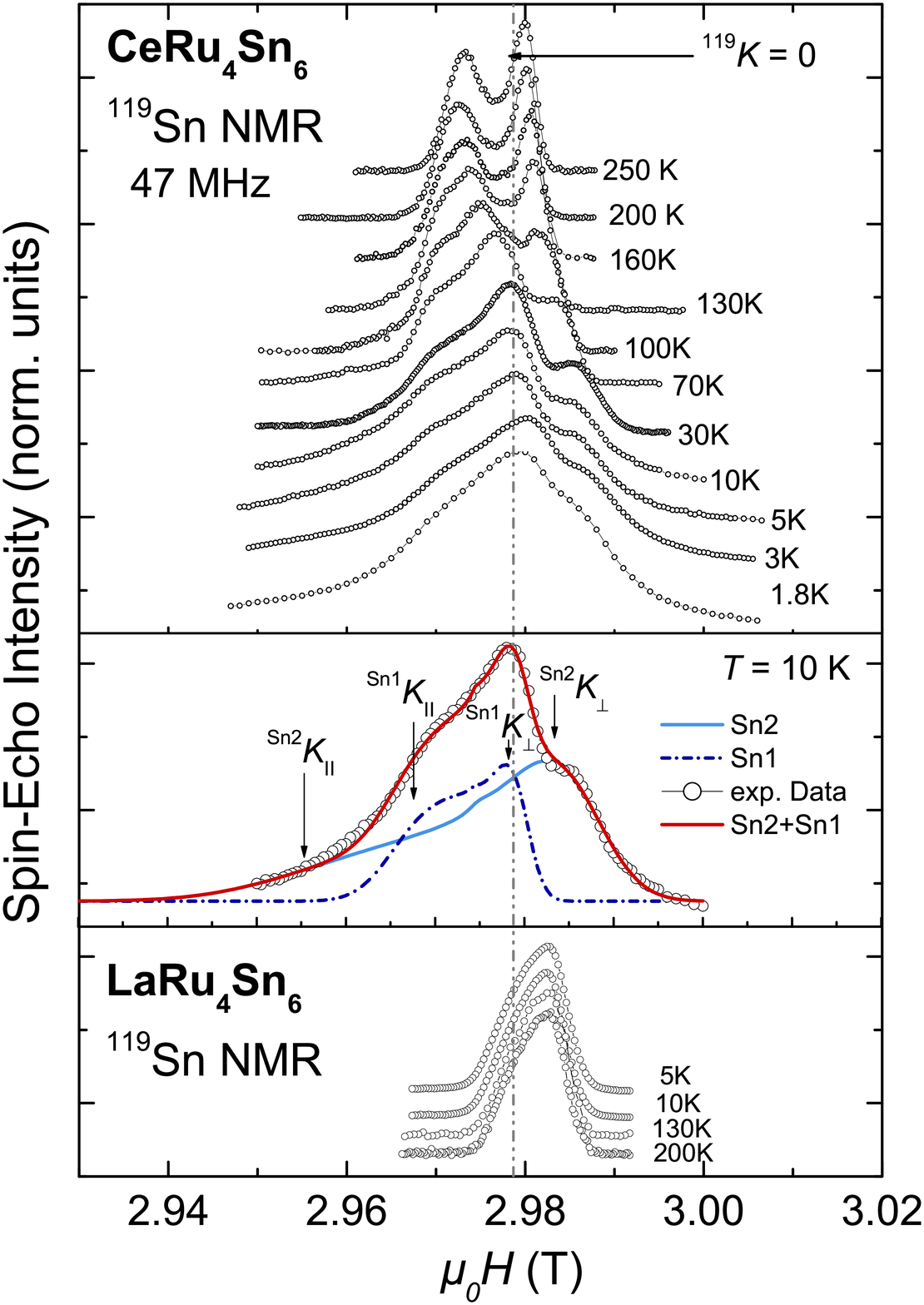}
\caption{(Color online) Field-dependent $^{119}$Sn-NMR spectra obtained
  at various sample temperatures. By comparison, the non-magnetic
  homologue \La\ produces an NMR response (bottom panel) with no
  discernable temperature variation. The middle panel shows a simulation
  of the spectra associated with the two inequivalent Sn-sites and the
  resulting spectrum; $K_{\perp}$ and $K_{\parallel}$ denote the local
  maxima attributed to the shift perpendicular and parallel to the
  $c$-axis, respectively.}
  \label{fig:spectra}
\end{figure}

To investigate the low-energy excitations, we have
chosen Sn-NMR as microscopic probe. Particularly, we selected the
$^{119}$Sn isotope (nuclear spin $I = 1/2$) which exhibits the highest
natural abundance (8.58\%) among the three isotopes. This allows for a
good NMR signal. In general, NMR can provide information about the
existence of an energy gap and its shape. In relation to studies on
Kondo insulators especially, detailed NMR studies on systems such as
CeNiSn~\cite{Kyogaku1990,NakamuraK1996},
Ce$_3$Bi$_4$Pt$_{3}$~\cite{Reyes1994}, SmB$_6$~\cite{Caldwell2007}, and
U$_2$Ru$_2$Sn~\cite{Rajarajan2007} have been shown to benefit from the
acute sensitivity of NMR response to changes in the charge-carrier
density of states and narrow gap formation. The spectra of the \Ce\
polycrystals were obtained with the field-sweep method at
\unit{47}{\mega\hertz} (Fig.~\ref{fig:spectra}, top panel). Due to the
two inequivalent sites occupied by Sn in the unit cell, the NMR spectrum
exhibits two lines with an ideal intensity ratio of 2:1. Therefore, the
powder pattern consists of a superposition of two anisotropic Sn1 and
Sn2 lines.
This is emphasized for a single spectrum at $\unit{10}{K}$ in
Fig.~\ref{fig:spectra} (middle panel). The well defined line shape
suggests that the sample is homogeneous and no disorder effects are
evident. To determine the Larmor field ($K_{\rm iso} = 0$), the
reference material $\alpha$-SnO with $^{119}K_{\rm iso} =
\unit{0.5}{\%}$ was used \cite{Carter1977} (dashed-dotted vertical line
in Fig.~\ref{fig:spectra}). It follows that the isotropic NMR shift of
the $^{119}$Sn spectra of \Ce\ is very small ($^{^{119}}K_{\rm iso}
\simeq \unit{0.5}{\%}$) and nearly temperature independent. In contrast,
the $^{119}$Sn-NMR spectra of the structurally related stannides
CeRuSn$_3$ and Ce$_3$Ru$_4$Sn$_{13}$, both of which are classified as
heavy-fermion metals \cite{Fukuhara1989,Takayanagi1990}, exhibit larger
and temperature-independent shifts ($^{119}K \simeq \unit{4}{\%}$) at
$\unit{4}{\kelvin}$. The small shift in \Ce\ reflects a weak hyperfine
coupling because of the low carrier density.

The spectra of \La\ show a Gaussian broadening and only a weak anisotropy (see
Fig.~\ref{fig:spectra}, bottom panel). \La\ is a good metal and a weak Pauli
paramagnet.
\begin{figure}[!ht] \centering
 \includegraphics[width=0.6\columnwidth]{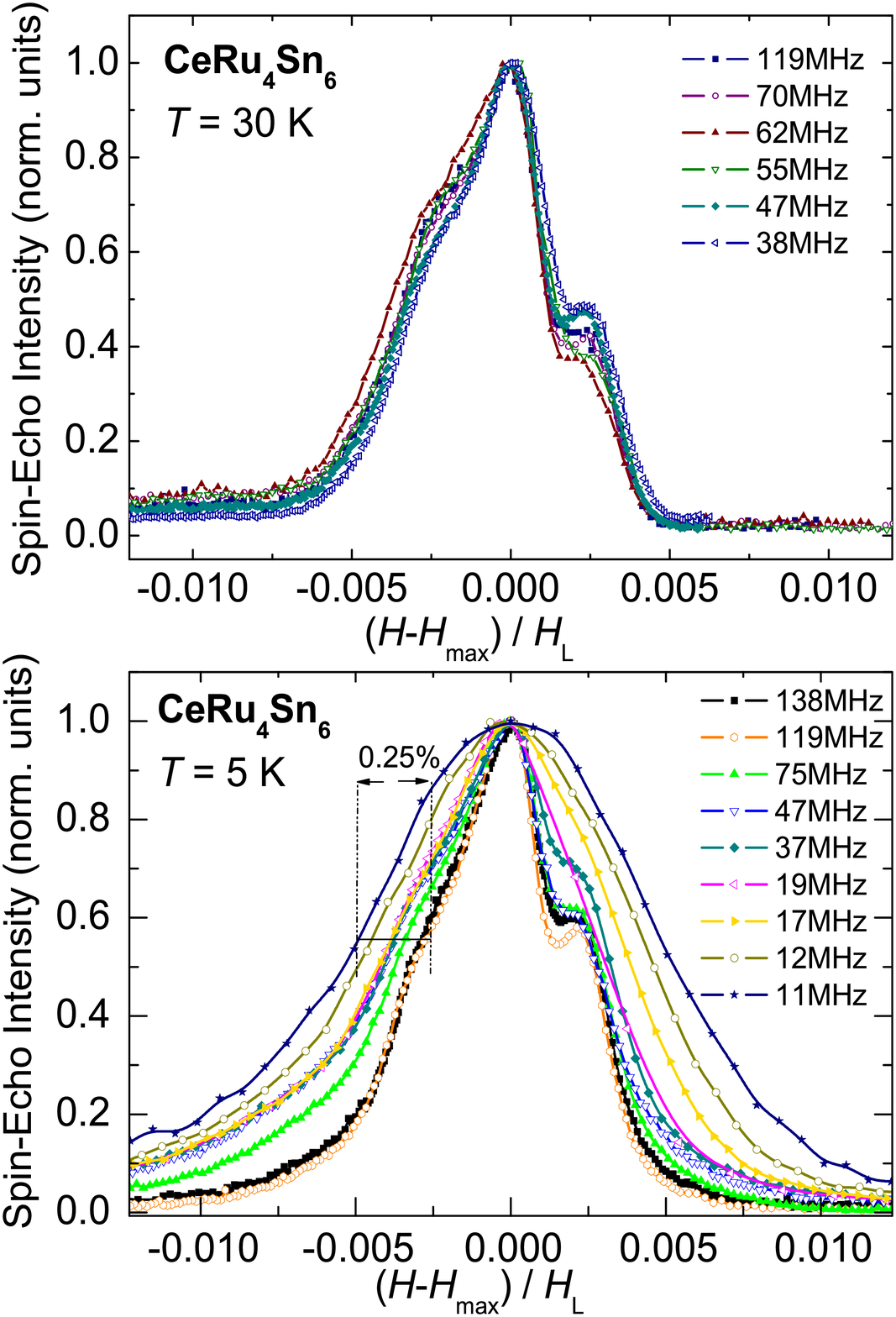}
  \caption{(Color online) $^{119}$Sn-NMR spectra at various frequencies
grouped by temperature to emphasize the line broadening occurring at low
temperatures. The comparatively small shift (see Fig.~\ref{fig:spectra})
has been normalized out.}
  \label{fig:Spec5K30K}
\end{figure}
A negative shift in this case might be attributed to a chemical shift.
At higher temperatures ($\unit{130}{\kelvin} \leq T \leq
\unit{150}{\kelvin}$), the $^{119}$Sn-NMR spectra of \Ce\ show two
distinct maxima which are attributed to the two Sn positions
$K_{\perp}$. Towards lower temperatures, the spectra broaden and the two
sharp maxima can no longer be easily resolved. The anisotropic shift of
the two Sn lines attributed to the spectral overlap of the two inequivalent
Sn sites imparts an arbitrary element to their simulation which makes a
quantitative analysis of these spectra rather tenuous. Therefore, we
limit our interpretation to what can be learnt from estimating the NMR
shift and the hyperfine field.

The respective frequency- and field-dependent investigation of the NMR
spectra, concerning shift and broadening, should give information about
the unusual behavior of the susceptibility at low temperatures
($T<\unit{10}{\kelvin}$). We measured field-sweep $^{119}$Sn-NMR spectra
at different frequencies and at two different temperatures, 30 and
\unit{5}{\kelvin}: The results are shown in Fig.~\ref{fig:Spec5K30K}. At
$T = \unit{30}{\kelvin}$, the distribution of frequencies is caused
entirely by the distribution of NMR shifts (Fig.~\ref{fig:Spec5K30K},
top panel) where field scans are plotted as shift distributions
$(H-H_{\rm max})/H$ for a range of NMR frequencies. The uniformity of
shape and broadening is seen to be quite precise. When scans taken at
\unit{5}{\kelvin} are plotted in the same fashion
(Fig.~\ref{fig:Spec5K30K}, bottom panel), a striking change is seen to
take place. A progressively larger (fractional) broadening occurs as the
frequency is lowered, so that at \unit{11}{\mega\hertz} the spectrum is
nearly twice as wide as at \unit{119}{\mega\hertz}. The analysis of the
entire series of spectra can be represented by a distribution of shifts
broadened by a convolution with a Gaussian function having a fixed width
parameter of $H = \unit{15}{Oe}$. Broadening fields which are
independent of the applied field only occur in cases of "spin freezing",
e.g. in a spin glass such as Cu:Mn \cite{Levitt1977}. Thus, the
field-independent broadening effect observed here corresponds to a
quasi-static component in the magnetic moments. CeRu$_4$Sn$_6$, on the
contrary, is not a spin glass: It does not have the kind of exchange
disorder that is found in spin glasses, and there is no cusp in the AC
susceptibility taken at small
fields. % (\textcolor{red}{who did this measurement?})
Additionally, very slowly fluctuating moments were detected by $\mu$SR
down to \unit{0.05}{\kelvin} \cite{Strydom2007}, without any "freezing"
of magnetic moments.

The spin-lattice-relaxation time was measured
with a saturation-recovery sequence at 47, 70, 100 and
\unit{119}{\mega\hertz}.
\begin{figure}[!ht] \centering
\includegraphics[width=0.97\columnwidth]{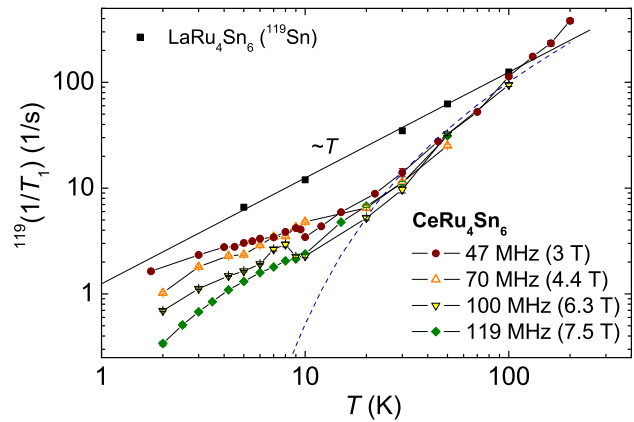}
  \caption{(Color online) Temperature evolution of the
spin-lattice-relaxation rate $^{119}(1/T_{1})$ of CeRu$_{4}$Sn$_{6}$ in
a double-logarithmic plot. (Dashed line: Fit of an exponential function, for
details see text). For comparison, the $1/T_{1} \propto T$ dependence of the reference \La\,
is also shown.}
  \label{fig:T1}
\end{figure}
The investigation was carried out at the spectral maximum which
corresponds to the $^{\rm Sn1}K_{\perp}$ position labeled in the powder
pattern plotted in Fig.~\ref{fig:spectra}. Due to the overlap of the NMR
signal of the two Sn positions, $T_1$ is composed of the contributions
attributed to the two Sn positions. Therefore, the recovery
magnetization $M(t)$ obtained from the measurements contains
contributions from both $H\perp c$ and $H \parallel c$ directions and
could not be described by a simple exponential behavior, $\propto
(1-\exp(-t/T_{1})$). Here, $M(t)$ could be fitted well by a stretched
exponential function:
\begin{equation}
  \label{eq:StrExp} M(t) = M_0 \cdot
\left[1-\exp\left(-\frac{t}{T_1}\right )^n\right]
\end{equation}
where $n = 0.5$ is a constant weighting factor. The data of
$^{119}(1/T_{1})$ extracted here should be treated as being powder
averaged. The spin-lattice-relaxation rate $^{119}(1/T_1)$ is shown as a
function of temperature in Fig.~\ref{fig:T1}. Below
$\unit{300}{\kelvin}$ down to $T \simeq \unit{30}{\kelvin}$, no
significant field nor frequency dependence is detected, in contrast to
the situation at lower temperatures where, $^{119}(1/T_1)$ decreases with
increasing field. $^{119}(1/T_1)$ of the non-magnetic homologue \La\ is plotted
in the same figure: The temperature dependence is linear as expected
from the Korringa relation $1/T_1 \propto T$ for a paramagnetic metal.
$^{119}(1/T_1)$ of \Ce\ deviates significantly from that of the
reference system compound \La\ towards lower temperatures. This is
characteristic of systems exhibiting the opening of an energy gap at the
Fermi level. This phenomenon is evident in many other KIs such as
U$_2$Ru$_2$Sn \cite{Rajarajan2007}, CeNiSn
\cite{NakamuraK1996,Kyogaku1990}, Ce$_{3}$Bi$_{4}$Pt$_{3}$
\cite{Reyes1994}, FeSb$_2$ \cite{Gippius2009} and SmB$_6$
\cite{Menth1969,Caldwell2007}. For non-cubic systems like CeNiSn and
U$_{2}$Ru$_{2}$Sn, a $T^{3}$ power law valid over two decades in
temperature was found in $^{119}(1/T_{1})$. A consistent description in
the entire temperature range could be done by applying the so called
$V-$shaped gap model for $N(E)$ \cite{NakamuraK1996,Rajarajan2007}. For
the cubic system Ce$_{3}$Bi$_{4}$Pt$_{3}$ a rectangular gap with a small
amount of in-gap states was used to describe $1/T_{1}$. For \Ce\ a
$T^{3}$ power law could not be found. Here, the existence of an energy
gap $\Delta E_{g1}$ can be verified quantitatively by fitting the
experimental data above approximately \unit{10}{\kelvin} with $1/T_1
\propto T N(E_F)^2 \propto T\exp(-\Delta E_{g1}/k_BT)$ according to the
Korringa relation, which gives in our case $\Delta E_{g1} =
\unit{30}{\kelvin}$ (dashed line in Fig.~\ref{fig:T1}). This value
is in good agreement with resistivity~\cite{Das1992} and
thermopower~\cite{Strydom2005} measurements. The deduced magnitude of
the energy gap seems not to be strongly dependent on the NMR magnetic
field and frequency used in these experiments. It is important to note
that the fit is found not to follow the $1/T_{1}$ results over the entire
temperature range, which is due to the crossover to a different power law
towards lower temperatures.

Below $T \simeq \unit{10}{\kelvin}$, the spin-lattice-relaxation rate
deviates from the exponential behavior and becomes strongly field
dependent. This field dependence of $T_{1}$ is a rare feature.
Nonetheless, similar results have been found for the $^{11}$B-NMR in the
Kondo insulator SmB$_6$ \cite{Caldwell2007}.
\begin{figure}[!ht] \centering
\includegraphics[width=0.97\columnwidth]{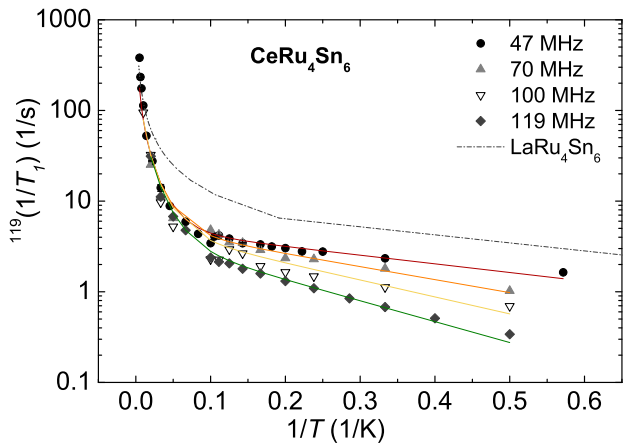}
\caption{(Color online) Arrhenius plot of the spin-lattice-relaxation
  rate to emphasize the two-gap origin of the temperature behavior (see
  equation~\eqref{eq:Eg1+Eg2}). The solid lines correspond to the
  results of a calculation in the frame of a two-gap model (for details see text).}
  \label{fig:CRST1Arrh}
\end{figure}
This has been attributed to low-energy excitations of magnetic states
within the energy gap at the Fermi level. The origin of these states is
still puzzling and controversial. In intermetallic compounds it is quite
natural to invoke the extrinsic involvement of chemical impurities as
opposed to more intrinsic origins. An objective view requires extrinsic
impurities to be considered as a possible origin for in-gap states.
Intrinsic donor states with a chemical origin and surface states derived
from the discontinuity of the bulk periodic potential are features
implicated even in the very clean limit of semiconducting charge
transport and may thus be deemed as intrinsic origins of in-gap states.
The comparatively narrow NMR line shape is indicative of a good sample
quality. Additionally, the $\mu$SR spectra \cite{Strydom2007} were found
to accurately follow simple, one-component exponential
time-decays, which are evidence of a homogeneous crystal environment
in the vicinity of the muon stopping site. Thus motivated by the
high quality sample material we took a closer look at the temperature
and field dependences of $1/T_{1}$ below \unit{10}{\kelvin}. Arranging
the data in an Arrhenius plot of $\log(1/T_{1})$ vs. $1/T$, we notice that
in the low-temperature region ($1/T > \unit{0.1}{\kelvin^{-1}}$),
$1/T_{1}$ is linear in $1/T$ for all frequencies. This range can be
fitted with $1/T_1(T) \propto \exp(-\Delta E_{g2}/k_BT)$, where the
small gap feature $\Delta E_{g2}/k_BT$ turns out to be strongly field
dependent (cf. table~\ref{tab:tab1}). Such a simple Arrhenius type of
exponential behavior is usually found in narrow spin-gap materials such
as organic spin-Peierls compounds \cite{Ehrenfreund1977}. The whole temperature range can be
well described by a two-gap equation:
\begin{equation}
  \label{eq:Eg1+Eg2} 1/T_1 = a \cdot T\cdot \exp(-\frac{\Delta
E_{g1}}{k_BT})+ b \cdot \exp(-\frac{\Delta E_{g2}}{k_BT})\,,
\end{equation}
with a fixed value of $\Delta E_{g1} = \unit{30}{\kelvin}$. The
quantities $a$ and $b$ are adjustable parameters. The nature of the results
obtained, i.e. $\Delta E_{g1} \simeq 10 \cdot \Delta E_{g2}$, imparts
confidence as to the additivity of the two exponentials in
Eq.~\eqref{eq:Eg1+Eg2}.
\begin{table}[!ht]
  \caption{\label{tab:tab1} Parameters resulting from fitting the
spin-lattice-relaxation rate versus temperature with
equation~\eqref{eq:Eg1+Eg2}.}
\begin{ruledtabular}
\begin{tabular}{ccc} $\nu$~(MHz) & $\mu_{0} H$~(T) & $\Delta
E_{g2}/k_B$~(K)\\ \hline 47 & 2.96 &2.23\\ 70 & 4.41 & 3.37\\ 100 & 6.30
& 4.86\\ 119 & 7.40 & 5.34\\

\end{tabular}
\end{ruledtabular}
\end{table}
The least-squares fits of the function in equation~\eqref{eq:Eg1+Eg2}
are shown as solid lines in Fig.~\ref{fig:CRST1Arrh}. This function
describes the experimental data quite well and provides a basis from
which to consider a two-gap scenario in CeRu$_{4}$Sn$_{6}$. The values
are summarised in table~\ref{tab:tab1}. This result suggests that a residual
density of states, $N(E_{F})$, exists at the Fermi level and develops an
energy gap $\Delta E_{g2}$ at $E_{F}$ which widens with increasing
magnetic field. Measurements other than NMR are supportive of this
description of the physics in CeRu$_{4}$Sn$_{6}$: (i) Resistivity data show a
saturation at low $T$, which suggest that a residual number of carriers
exist and participate in the electronic transport \cite{Das1992}; (ii) in the
thermopower measurements, a maximum in $S(T)$ is found at $T \simeq
\unit{3}{\kelvin}$ (zero field) which is shifted at \unit{9}{\tesla} to
$T \simeq \unit{5}{\kelvin}$ (Fig.~\ref{fig:thermo}); % achieved in an applied
% field at $T \simeq \unit{5}{\kelvin}$ whereas this is absent in zero
% field measurements;
(iii) the Sommerfeld coefficient shows a maximum in zero field which
moves towards higher temperatures with field as well
(Fig.~\ref{fig:CRSCpT}); and (iv) the Sommerfeld coefficient in zero
field is strongly enhanced and indicates the existence of electronic quasiparticles
with large effective masses.

\subsection{Low-temperature specific heat and narrow-band model}

\label{sec:model} Since the specific heat is directly linked to the
density of states at the Fermi level through

\begin{equation}
  \label{eq:SpH} C = \frac{dU}{dT} = \int_{-\infty}^{+\infty} E N(E)
\frac{df(E)}{dT}dE~,
\end{equation}

\noindent where $U$ is the free energy and $f(E)$ the
Fermi function \cite{Gopal1966}, we have analysed $C(T)/T$ assuming for
the density of states two narrow bands represented by a Cauchy-Lorentz
function:

\begin{equation}
  \label{eq:CL} N(E) = z\cdot\frac{1}{2\pi}\left[
\frac{\Gamma}{(E-\Delta)^{2}+\Gamma^{2}}+
\frac{\Gamma}{(E+\Delta)^{2}+\Gamma^{2}}\right]~.
\end{equation}

\begin{figure}[!ht] \centering
  \includegraphics[width=0.97\columnwidth]{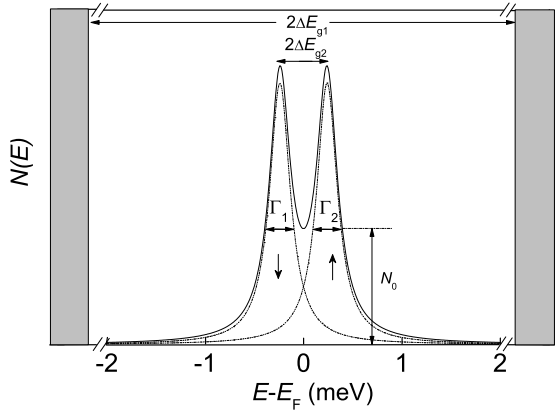} %\includegraphics[width=10cm]{Fig10.eps}
  \caption{Model of the residual field-dependent states in the large
    hybridization gap $\Delta E_{g1}$. This model was used for the
    calculation of $C(T,H)$ and $T_{1}^{-1}(T,H)$ of CeRu$_{4}$Sn$_{6}$.
  Please note that the free parameters $\Delta$, $\Gamma$ and $z$ are
  field dependent and temperature independent. Values used for the
  calculation are listed in Table~\ref{tab:table2}; Fig.~\ref{fig:DOSfu}
  shows the resulting field-dependent plots of $N(E)$.}
  \label{fig:DOSE}
\end{figure}

The quantity $\Delta$ denotes the energy gap $\Delta E_{g2}$, $\Gamma$ is the full width
at half maximum and the parameter $z$ is a weighting factor. A plot of
this function is illustrated in Fig.~\ref{fig:DOSE}.

\begin{figure}[!ht] \centering
\includegraphics[width=0.97\columnwidth]{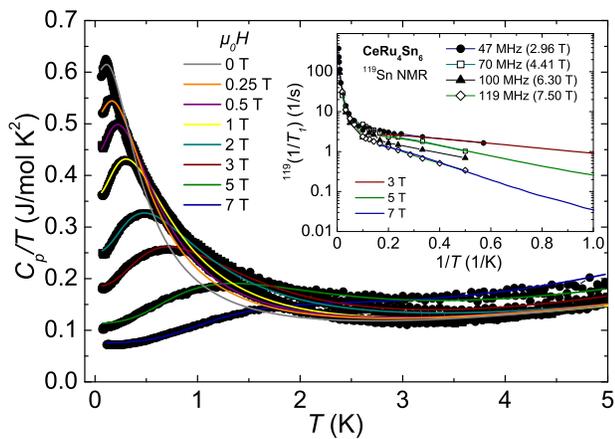}
  \caption{(Color online) Temperature dependence of the specific heat (main panel) and
spin-lattice-relaxation rate (inset) together with fits for each applied
field, according to Eq.~\eqref{eq:SpH}, \eqref{eq:CL} and \eqref{eq:3}, respectively.}
  \label{fig:SpH+NMR}
\end{figure}
The density of states considered here represents only the
in-gap states by means of the trial function since the real density of
states cannot be determined exactly with the data at hand. In the
following, we discuss the application of this function and the adherence
of the calculated curve to the measured data.

\begin{figure}[t] \centering
\includegraphics[width=0.97\columnwidth]{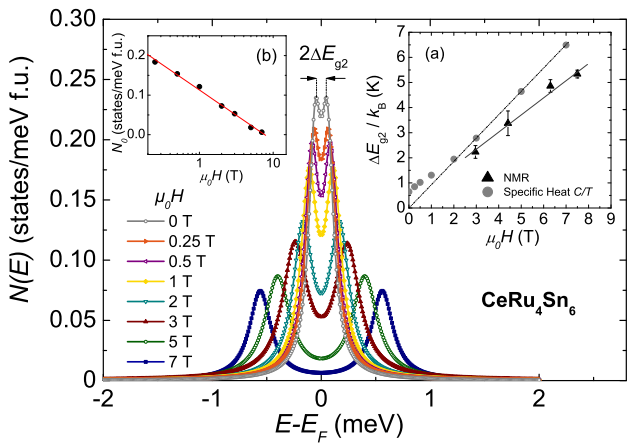}
  \caption{(Color online) Density of states projection by means of
equation~\eqref{eq:CL} against energy relative to the Fermi energy
$E_{F}$. The field-derived $T \to 0$ states density is depicted
in inset (b) and inset (a) plots the field evolution of the inner gap
$\Delta E_{g2}$ responsible for the low-energy properties. Lines are
guides to the eye.}
  \label{fig:DOSfu}
\end{figure}

It is shown in Fig.~\ref{fig:SpH+NMR} that a gapped band structure and
the density-of-states enumeration accurately describes specific-heat and
NMR results in a consistent manner. To fit the specific-heat data at all
fields with just three parameters ($\Delta$, $\Gamma$ and $z$), the
nuclear Schottky contribution was subtracted and the high-temperature
range was considered field independent with a Sommerfeld coefficient of
\unit{0.08}{\joule\per\mole\kelvin\squared} and a phonon contribution
with a Debye temperature of $\Theta_{D} = \unit{250}{\kelvin}$. The
theoretical description accurately accounts for the $T$- and $H$-
dependent specific-heat data. The parameters used to fit the data are
listed in Table \ref{tab:table2}, and a graphical representation is
illustrated in Fig.~\ref{fig:DOSfu}.
\begin{table}[!ht]
  \caption{\label{tab:table2} List of the parameters used to fit
$C(T)/T$ vs. $T$ with the density of states of the form given in
equation~\ref{eq:CL}. The $X^{2}$ parameter indicates the adherence
of the fits.}

\begin{ruledtabular}
\begin{tabular}{ccccc} $\mu_{0}H$~(T) & $\Delta$~(meV) &
$\Gamma$~(meV) & $z$ &
$X^{2}$~($10^{-5}$)\\ \hline
0 & 0.056 & 0.068 & 12.8 & 20\\
0.25 & 0.073 & 0.079 & 11.9 & 7 \\
0.5 & 0.088 & 0.080 & 11.8 & 5.7 \\
1 & 0.113 & 0.090 & 11.4 & 5\\
2 & 0.168 & 0.110 & 12.0 & 5 \\
3 & 0.240 & 0.138 & 10.9 & 2 \\
5 & 0.400 & 0.140 & 13.5 & 3 \\
7 & 0.560 & 0.120 & 18 &  2 \\

\end{tabular}
\end{ruledtabular}

\end{table}
To check the validity of the model, we have used the same $N(E)$
functions to calculate the spin-lattice-relaxation rate, because it can be
described by \cite{Riseborough2000}:
\begin{equation}
  \label{eq:3} 1/T_{1} \propto H^{2}_{\rm hf} \int N(E)^{2}
f(E)[1-f(E)]dE
\end{equation}
where the only adjustable parameter is the hyperfine field $H_{\rm hf}$, the
role of which is only to shift up and down the fitting curves of an
Arrhenius plot. The slope of these lines is given by the size of
$\Delta$. The results are shown in the inset of Fig.~\ref{fig:SpH+NMR}
for three different magnetic fields, and these fits describe the
experimental data very reliably.

The change of the density of states with increasing magnetic field is
shown in Fig.~\ref{fig:DOSfu} and the behavior reminds of the $s= 1/2$
Kondo imputiy model \cite{Costi2000}. As a matter of fact, we can now use this
density of states to calculate all the observables. The point $N_{0}$
where $N(E)$ crosses the Fermi level provides the Sommerfeld coefficient
$\gamma(0)$ at $T=0$. Plotting $N_{0}$ as a function of $\log(\mu_{0}H)$
indicates how $\gamma$ decreases in an exponential fashion with
increasing magnetic field (see inset (a) of Fig.~\ref{fig:DOSfu}). To
compare the values of $\Delta E_{g2}/k_{B}T$ obtained from NMR
measurements and those obtained by fitting the specific-heat data, we
plotted both in the same diagram in inset (b) of the same figure.
Similar values and a comparable field dependency is found among the two
approaches. Assuming a Zeeman energy $\Delta E = -\vec{\mu} \cdot
\vec{B}$, the magnetic moment of these states can be estimated from the
slope of the dashed line in inset (b) of Fig.~\ref{fig:DOSfu}: The fit
gives a value close to $\unit{1}{\mu_{B}}$.
To estimate the number of states per formula unit, $N(E)$ can be
integrated. At zero field and $T = \unit{2}{\kelvin}$, 0.039 states per
formula unit are obtained. This value is in good accordance with that
obtained from Hall-coefficient data, which gave 0.03 carriers
per formula unit~\cite{Strydom2005}. Additionally, the contribution of
these spin states to the magnetization can be calculated at given values of
temperature and field. Considering that the states have $s=1/2$, the
magnetization can be calculated by
\begin{equation}
  \label{eq:Magn} M = \mu_{B}(n_\uparrow - n_\downarrow)
\end{equation}
where
\begin{equation}
  \label{eq:n-up} n_\uparrow = \frac{1}{2} \int_{-\infty}^{+\infty}
N(E+\mu_BH)f(E)dE
\end{equation}
and
\begin{equation}
  \label{eq:n-down} n_\downarrow = \frac{1}{2} \int_{-\infty}^{+\infty}
N(E-\mu_BH)f(E)dE\,.
\end{equation}
\begin{figure}[!ht] \centering
\includegraphics[width=0.97\columnwidth]{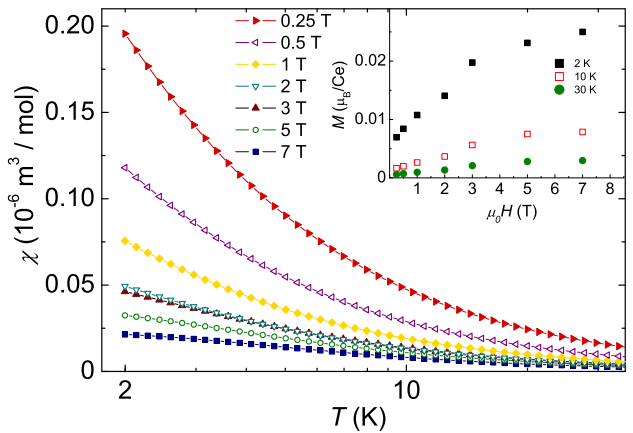}
  \caption{Calculated susceptibility (main panel) and magnetization
(inset), guided by the density-of-states calculation of
Eq.~\eqref{eq:CL}.}
  \label{fig:MagCalc}
\end{figure}
For example, at $\mu_{0}H = \unit{7}{\tesla}$ and $T =
\unit{2}{\kelvin}$, $(n_{\uparrow}-n_\downarrow)\simeq 0.027$ states are
obtained, which cause a net magnetization $M \simeq
\unit{0.027}{\mu_B}$/f.u. The experimental measurements
(Fig.~\ref{fig:Magn}) give $M \simeq \unit{0.14}{\mu_B}$/f.u. at the
same values of temperature and field; this means that only 20\% of the
residual density of states contribute to the complete magnetization,
where the van-Vleck contribution to $M$ has, however, not been taken
into account. The field dependence of the magnetization due to the
in-gap states only is shown in the inset of Fig.~\ref{fig:MagCalc} and
is found very similar to the total magnetization in
Fig~\ref{fig:Magn}.

The model for the density of states (Eq.~(\ref{eq:CL})) could also
be successfully applied to calculate the complete temperature and field
dependences of the susceptibility. This calculation for selected fields
is illustrated in Fig.~\ref{fig:MagCalc}. Qualitatively, the steep
increase of $\chi(T)$ at low temperatures can be reproduced by such a
form of the residual density of states. Below \unit{2}{\kelvin},
$\chi(T)$ flattens and its value $\chi$ at $T=0$ decreases with
increasing field, scaling with $\gamma(0)$. The predicted trend is
largely played out by the experimental results in Fig.~\ref{fig:susc}.
However, the absolute values differ generally from the experimental
ones. It should be noted that additional effects such as an underlying
magnetocrystalline anisotropy precludes a quantitative treatment in the
present study on polycrystalline samples.
%
%---------------------------------------Summary-------------------------------------------
%
\section{Summary and Conclusions}
\label{sec:summary}

We have presented detailed measurements of the field response of the
specific-heat ($C$) and of nuclear magnetic resonance (NMR) on the
intermetallic Kondo-insulator compound CeRu$_4$Sn$_6$. A survey of
earlier results of the electrical resistivity, thermopower, and
specific-heat hinted at the possibility that this system might be on the
verge of magnetic order, and that magnetic correlations certainly play
an influential role in the anomalous ground-state properties. Here,
strong correlations form out of a low charge carrier density which is a
rare occurrence in the highly correlated class of systems. The aim of
this work was to seek a description that would consistently explain the
strongly enhanced low-temperature specific heat and magnetic phenomena
found in \Ce\ through NMR measurements. A V-shaped pseudogap with a
broad band of residual states at the Fermi energy was first proposed by
Kyogaku and coworkers \cite{Kyogaku1990} and provides a convenient
mechanism for modeling susceptibility, Knight shift, and
spin-lattice-relaxation rate in non-cubic Kondo insulators like CeNiSn
and U$_{2}$Ru$_{2}$Sn. For CeRu$_{4}$Sn$_{6}$, this model is not valid,
and we have assumed a simple, rigid rectangular gap model and have
replaced the broad band by a narrow one of (field dependent) residual
states. We have applied this density-of-states model with considerable
success to the specific-heat and NMR data of \Ce\ and found the field
dependence of these properties to be fully reconcilable in the framework
of this model. The temperature evolution of electronic and thermodynamic
properties is driven by two nested energy gaps, $\Delta E_{g1} =
\unit{30}{\kelvin} $ and $\Delta E_{g2}\approx \unit{0.65}{\kelvin}$ (at
zero field), centered on the Fermi energy. The gapping is not complete
and a residual number of heavy-mass charge carriers within the smaller
of the two gaps achieves a finite electrical conductivity even at the
lowest temperatures. These states are correlated, presumably due to the
Kondo-screening effect, as inferred from strong temperature dependences
of the Sommerfeld-coefficient of 4$f$ increment to the specific heat and
the magnetic susceptibility, both of which are found to saturate at
large values as $T \to 0$.

The scope of the physics forwarded for \Ce\ in this work provides a
platform from which to test the predicted nodal Kondo insulating state
as a new type of semimetal that forms under favorable conditions
\cite{moreno2000}. The axial symmetry of nodes of vanishing density of
states set the stage for anisotropic electronic conduction. Anisotropic
magnetic properties in \Ce\ single crystals has recently been reported
\cite{Paschen2010}. A considerable magnetocrystalline anisotropy is
evident between the basal-plane and the tetragonal $c$-axis. In
particular, a tiny $c$-axis magnetization of $\unit{0.065}{\mu_B}$ at
$\unit{3}{\kelvin}$ was found in $\unit{6}{\tesla}$ which amounts to no
more than about $20\%$ of the moment extracted along the isotropic basal
plane directions. Furthermore, our results obtained might reflect the
interplay between  RKKY and Kondo interaction in the limit of a very low
carrier concentration as discussed by Coqblin \emph{et al.}
\cite{Coqblin2003}. % This is the rare case where the number of
% conduction electrons is not large enough to screen all spins.

Future studies on \Ce\ would benefit from magnetic susceptibility data
at very low temperatures to assess, for instance, the role of magnetic
correlations through progression of the Sommerfeld-Wilson ratio.
Electrical resistivity studies on single crystals are highly desirable and
high-resolution inelastic-neutron-scattering measurements would
conceivably enhance our knowledge about the low-lying spin excitations
in the material.

%----------------------Acknowledgement------------------------------------
\section{Acknowledgement}
\label{sec:acknowledgement} The authors thank P.\
Coleman, D.\ Adroja, S.\ Paschen, and O.\ Stockert for stimulating
discussions. AMS thanks the SA-NRF (Grant 2072956) and the University of
Johannesburg Research Committee.
% ------------------------------------------------------------------------

%--------------------Bibliography------------------------------------------

%-------------------------------------------------------------------------
\end{document}